\documentclass[12pt]{article}
\usepackage[bottom]{footmisc}
\usepackage[utf8]{inputenc}
\usepackage{authblk}
\usepackage{epsfig}
\usepackage{amssymb}
\usepackage{amsfonts}
\usepackage[mathscr]{euscript}
\usepackage{amsbsy}
\usepackage[all]{xy}
\usepackage{amsmath}
\usepackage{enumerate}
\usepackage{float}

 \usepackage{enumerate}
 \usepackage{hyperref}

\usepackage{amssymb,amscd}
\usepackage{amsmath,amsthm}
\usepackage{xspace}
\usepackage{url}
\newcommand{\del}{\partial}
\newcommand{\ov}{\overline}

\usepackage{epsfig}
\usepackage{amssymb}
\usepackage{amsfonts}
\usepackage{amsbsy}
\usepackage[all]{xy}
\usepackage{amsmath}
\usepackage{enumerate}

\usepackage{amssymb,amscd}
\usepackage{mathrsfs}
\usepackage{amsmath,amsthm}
\usepackage{xspace}
\usepackage{url}

\newcommand{\bea}{\begin{eqnarray}\displaystyle}
\newcommand{\eea}{\end{eqnarray}}

\title{On the $A_{\infty}$-Category of a Holomorphic Moment Map}
\author{Ahsan Z. Khan\footnote{{\tt khan@ias.edu}}}
\affil{School of Natural Sciences \\ Institute for Advanced Study \\ Einstein Drive, Princeton NJ 08540 }

\begin{document}

\maketitle 

\begin{abstract}
Let $M$ be a hyperK\"{a}hler manifold equipped with a $U(1)$ hyperK\"{a}hler isometry, and let $I$ be a complex structure on $M$. In this note, we study the $A_{\infty}$-category of A-branes for the Landau-Ginzburg model with target space $(M,I)$, and superpotential being the $I$-holomorphic moment map. We show that if $I$ is a generic complex structure, the $A_{\infty}$-category is semi-simple. For exceptional complex structures, though typically not semi-simple, the category still has no instanton corrections. We illustrate the $A_{\infty}$-category at both generic and exceptional loci when $M$ is the cotangent bundle of the projective line. 
\end{abstract}

\paragraph{} The Morse-Smale-Witten (MSW) complex is a cochain complex associated to a real function\footnote{Throughout this paper we assume that all critical points of $h$ are isolated and non-degenerate} $h$ on a Riemannian manifold $(M,g)$. From a physics perspective, the MSW complex is the space of perturbative ground states of an $\mathcal{N}=2$ supersymmetric quantum mechanics (SQM) system \cite{Witten:1982im}. The differential on this complex is constructed from instanton effects.

\paragraph{} There is a special situation in which the MSW complex simplifies rather dramatically. This is when $(M,g)$ admits a $g$-compatible complex structure $I$ (so that $(M,g,I)$ is a K\"{a}hler manifold), and $h$ is the real part of an $I$-holomorphic function $W$ on $M$. If this is the case, it is well-known that all critical points of $h$ have the same Morse index, equal to the complex dimension of $M$. The MSW complex is thus simply a vector space concentrated in a single degree, with one basis vector for each critical point. This immediately implies that the differential must vanish.  In terms of physics, there is also something special that happens in this situation. The $\mathcal{N}=2$ supersymmetry of the SQM with target space $(M,g)$ and superpotential $h$ enhances to $\mathcal{N}=4$ precisely when $M$ is K\"{a}hler and $h= \text{Re} \,W$. Systems with twice the supersymmetry often have simpler properties for their supersymmetric ground states. 

\paragraph{Remark} Another case where the supersymmetry enhances to $\mathcal{N}=4$ is when $h$ is the moment map of a continuous K\"{a}hler isometry of $M$. We will use a complex version of this statement later in the note. 

\paragraph{} We conclude that the MSW complex of $\text{Re} \,W$ is not a very interesting invariant of the pair $ \big((M,g,I), W \big)$. There is, however, a richer invariant of the same pair which, in a sense that can be made precise, categorifies the MSW complex of $\text{Re}\,W$. The categorification of the MSW complex of $\text{Re}\,W$ is known (to mathematicians) as the Fukaya-Seidel $A_{\infty}$-category of $(M,W)$. In terms of physics what's responsible for the existence of this categorification is the fact that there is an uplift of the $\mathcal{N}=4$ SQM to a two-dimensional $\mathcal{N}=(2,2)$ Landau-Ginzburg (LG) model such that the original SQM supercharge lifts to a topological supercharge\footnote{A supercharge $Q$ is called topological if all translations are in its image.} of the LG model. The Fukaya-Seidel $A_{\infty}$-category is then simply the category of boundary conditions of the Landau-Ginzburg model that preserve this topological supercharge. These boundary conditions are known as A-branes. The MSW complex of $\text{Re} \,W$ can be recovered upon taking the Hochschild homology of the Fukaya-Seidel category. 

\paragraph{} To be more precise, there are a few different versions of the $A_{\infty}$-category associated to $(M,W)$ (all conjectured to be $A_{\infty}$-equivalent). Usually the term ``Fukaya-Seidel category" refers to a formulation by Seidel based on vanishing cycles and their intersections \cite{Seidel}. We will be working mostly with a formulation developed by Gaiotto, Moore and Witten, which is based on particular kinds of solutions to the $\zeta$-soliton and $\zeta$-instanton equations\footnote{A similar (but not identical) formulation was also developed by Haydys \cite{Haydys:2010dv}.} \footnote{See also \cite{Kapranov:2014uwa} for a reformulation of the formalism of \cite{Gaiotto:2015aoa} that generalizes to higher dimensions.} \cite{Gaiotto:2015aoa}. 

\paragraph{} It is natural to wonder if there is a class of K\"{a}hler manifolds and superpotentials for which the associated $A_{\infty}$-category simplifies. A possible criteria for such pairs could be that the supersymmetry of the Landau-Ginzburg model of $(M,W)$ enhances. There is indeed a special situation where this is the case. Suppose that the target space $(M,g,I)$ admits an $I$-holomorphic two-form $\Omega$ satisfying  \bea \Omega g^{-1} \ov{\Omega} + g =0, \,\,\,\,\, \nabla \Omega = 0, \,\,\,\,\,\,\, \eea  so that the space $(M,g,I, \Omega)$ is a hyperK\"{a}hler manifold\footnote{Recall that one definition of a hyperK\"{a}hler is as a K\"{a}hler manifold with a compatible parallel holomorphic symplectic form $\Omega$.}. Moreover, suppose $(M,g,I, \Omega)$ admits a hyperK\"{a}hler isometry generated by a vector field $V$, so that \bea \mathcal{L}_V g = \mathcal{L}_V I = \mathcal{L}_V \Omega = 0, \eea  and $W$ is the corresponding holomorphic moment map \bea \text{d}W = \iota_{V} \Omega.\eea If this is the case, the Landau-Ginzburg model for the pair $\big((M,g,I), W \big)$ has an additional symmetry\footnote{The axial R-symmetry $F_A$, the symmetry coming from rotating by $J$, and their commutator, generate an $\text{so}(3)$ R-symmetry algebra. Another way of seeing both the supersymmetry enhancement and this $SO(3)$ R-symmetry is to note that the two-dimensional theory comes from reducing a six dimensional hyperK\"{a}hler sigma model, where the hyperK\"{a}hler isometry is gauged by a six-dimensional $U(1)$ vector multiplet, to two dimensions. When doing the reduction, we set the gauge fields in the four internal directions to a non-zero vector in $\mathbb{R}^4$, while setting all other vector multiplet fields to vanish. The choice of a non-zero vector breaks the $SO(4)$ R-symmetry coming from the internal directions to $SO(3)$.} coming from rotating the fermions of the theory by the endomorphism \bea J = g^{-1}\text{Re} \big(\Omega \big)\eea of the tangent bundle. The commutator of this symmetry transformation with the $\mathcal{N}=(2,2)$ supersymmetries generates four additional fermionic symmetries resulting in a total of eight supercharges. The supersymmetry of the Landau-Ginzburg model of $(M, W)$ therefore enhances from $\mathcal{N}=(2,2)$ to $\mathcal{N}=(4,4)$.

\paragraph{} The above example in fact gives us an infinite family of pairs of K\"{a}hler manifolds and holomorphic functions for which the supersymmetry enhances. In addition to $I$ we can define the integrable complex structures $J = g^{-1} \text{Re}(\Omega)$ and $K= g^{-1} \text{Im}(\Omega)$ that moreover satisfy the quaternion relations \bea I^2 = J^2 = K^2 = IJK = -1.\eea This brings us to the well-known fact that a hyperK\"{a}hler manifold $M$ has a two-sphere's worth of complex structures, since for any $v=(v_1, v_2, v_3) \in S^2$, the endomorphism \bea I^{(v)} = v_1 I + v_2 J + v_3 K \eea is an integrable complex structure on $M$. Moreover, given a hyperK\"{a}hler isometry generated by a vector field $V$, there is a holomorphic moment map $$W^{(v)}: M \rightarrow \mathbb{C}$$ in every complex structure $I^{(v)}$. Explicitly, $W^{(v)}$ can be obtained as follows. Let $\mu_I$ be the moment map corresponding to the real symplectic form, $\omega_I = gI$ and let \bea W = \mu_J + \text{i} \mu_K , \,\,\,\,\, \Omega = \omega_J + \text{i} \omega_K, \eea so that the triple $(\mu_I, \mu_J, \mu_K)$ satisfies \bea \text{d}\mu_I = \iota_V \omega_I, \\\text{d} \mu_J = \iota_V \omega_J,\\ \text{d} \mu_K = \iota_V \omega_K.\eea We can obtain the complex structure $I^{(v)}$ from $I$ by a hyperK\"{a}hler rotation: there is a quaternion \bea q= q_0 + q_1 i + q_2 j + q_3 k \in \mathbb{H} \eea such that \bea \bar{q} \, i \,q = v_1 i + v_2 j + v_3 k.\eea The quaternion $q$ is unique up to a redefinition by a phase $q \rightarrow e^{\text{i}\theta}q$. Organize the hyperK\"{a}hler moment map in terms of the imaginary quaternions \bea \vec{\mu} = \mu_I  i + \mu_J j + \mu_K k. \eea With respect to the complex structure $I$, the decomposition of the hyperK\"{a}hler moment map into real and complex parts is given by \bea \vec{\mu} = h \,i + W j,\eea so that $h = \mu_I$ is the real moment map and $W = \mu_J + \text{i} \mu_K$ is the complex moment map.  The real and holomorphic moment maps with respect to $I^{(v)}$ are then given by doing a rotation of $\mu$ with respect to $\ov{q}$: \bea q \, \vec{\mu} \,\ov{ q}= h^{(v)} i + W^{(v)} j.  \eea Redefinition of $q$ by a phase, \bea q \rightarrow e^{\text{i} \theta}q \eea leaves $h^{(v)}$ invariant whereas it transforms \bea W^{(v)} \rightarrow e^{2\text{i}\theta}W^{(v)}. \eea Since redefinition of $W^{(v)}$ by a phase does not change the associated Landau-Ginzburg model, we get an LG model $((M, g,I^{(v)}), W^{(v)})$ for every point on the unit two-sphere.  Because the supersymmetry enhances for all such models, it is natural to expect that the $A_{\infty}$-category of each such pair $\big((M,g,I^{(v)}), W^{(v)} \big)$ simplifies.

\paragraph{} The purpose of this note is to show that this is indeed the case\footnote{For previous work that is related to the setting of the present note see \cite{Solomon:2018kww} and \cite{jin}}. 

\paragraph{} For the rest of the note we assume that the hyperK\"{a}hler moment map has isolated and non-degenerate critical points, and $v \in S^2$ is such that the critical values of $W^{(v)}$ are in general position on the complex plane. 

\paragraph{} The first basic simplification in the $A_{\infty}$-category of a moment map is the lack of instanton corrections. This can be seen as follows. Recall that in the Gaiotto-Moore-Witten formulation of the $A_{\infty}$-category of $W$, all instanton corrections come from solutions of the $\zeta$-instanton equation. The $\zeta$-instanton equation for a map $\phi: \mathbb{C} \rightarrow M$ is \bea \frac{\del \phi^i}{\del \bar{z}} = \zeta g^{i\bar{j}} \frac{\del \ov{W}}{\del \bar{\phi}^{\bar{j}}},\eea where $\zeta = e^{\text{i}\theta}$ is a phase we are required to choose in order to define the category, $(\phi^i, \ov{\phi}^{\bar{i}})$ are local complex coordinates on $M$ and $z$ is the standard complex coordinate on $\mathbb{C}$. The particular solutions that contribute to the $A_{\infty}$-structure are solutions with ``fan-like" boundary conditions at infinity, that are \underline{rigid}.  Rigidity means that a solution has no moduli other than overall translations of the Euclidean spacetime complex plane. That there cannot be any such solution in the hyperK\"{a}hler case can be easily seen from the fact that the operator obtained from linearizing the $\zeta$-instanton equation has a zero-mode \bea \delta \phi^i  = V^i \eea in addition to the translational zero-mode. There are therefore no non-trivial rigid instantons.

\paragraph{} Next we prove that at a generic point $v$ on the two-sphere of complex structures, a stronger statement holds: the $A_{\infty}$-category of the pair $ \big((M,g,I^{(v)}), W^{(v)} \big)$ is semi-simple. What we mean by this is the following. Recall that $A_{\infty}$-category of a superpotential $W$ has a generating set of ``thimble" objects $\{T_i\}$ that are in one-to-one correspondence with critical points of $W$. The $A_{\infty}$-category is said to be semi-simple if the morphism spaces between the thimble objects satisfy \bea \text{Hom}(T_i, T_j) = \delta_{ij} \mathbb{C}.\eea This property follows from an elementary Lemma.

\paragraph{Lemma} Let $(M, g, I,J,K)$ be a hyperK\"{a}hler manifold, $$ \vec{\mu}: M \rightarrow \mathbb{R}^3 $$ be a hyperK\"{a}hler moment map, and $\vec{n} = (n_1, n_2, n_3) \in S^2$ be a point on the unit sphere. Suppose $\phi: \mathbb{R} \rightarrow M$ is a (non-constant) solution to the gradient flow equation \bea \frac{\text{d}\phi^A}{\text{d}y} = g^{AB} \frac{\del  (\vec{n} \cdot \vec{\mu})}{ \del \phi^B}\eea where \bea \vec{n} \cdot \vec{\mu} = n_1 \mu_I + n_2 \mu_J + n_3 \mu_K.\eea Then the composition $\vec{\mu} \circ \phi: \mathbb{R} \rightarrow \mathbb{R}^3$ is an embedding with image a straight line in the $\vec{n}$-direction. 

\begin{proof} We show the claim for $\vec{n}=(0,1,0)$, so that we study the gradient flow equation for $\mu_J$. Since $\mu_J$ is the real part of the $I$-holomorphic function \bea W_I = \mu_J + \text{i} \mu_K,\eea the Cauchy-Riemann equation \bea \text{d} \mu_J = I^{\text{t}} \text{d}\mu_K \eea implies that the the gradient vector field of $\mu_J$ is equivalent to the Hamiltonian vector field of $\mu_K$ with respect to the symplectic form $\omega_I$ \bea g^{-1} \text{d}\mu_J = \omega_I^{-1} \text{d}\mu_K.\eea Therefore $\mu_K$ is constant along a flow line. But $\mu_J$ is also the real part of the $K$-holomorphic function $$W_K = \mu_J - \text{i} \mu_I$$ and so the Cauchy-Riemann equation for $W_K$ implies that the gradient flow of $\mu_J$ is also equivalent to the Hamiltonian flow for $-\mu_I$ with respect to the symplectic form $\omega_K$. Therefore $\mu_I$ is also constant. Since $\mu_J \circ \phi$ is monotonic, the image of the $\vec{\mu} \circ \phi$ is a line parallel to the $J$-axis. The case when $\vec{n}$ is arbitrary can be obtained by a hyperK\"{a}hler rotation. \end{proof}

\paragraph{} Consider now the Landau-Ginzburg model with target space $(M, g, I^{(v)})$ and superpotential given by the $I^{(v)}$-holomorphic function $W^{(v)}$. A fundamental role in Landau-Ginzburg models is played by BPS solitons. Let $\phi_i$ and $\phi_j$ be critical points of $W^{(v)}$. Recall that an $ij$-BPS soliton is a solution of the gradient flow equation for $\text{Re}(e^{-\text{i} \theta_{ij}} W^{(v)})$ where \bea e^{\text{i} \theta_{ij}} = \frac{W^{(v)}(\phi_i) - W^{(v)}(\phi_j)}{|W^{(v)}(\phi_i) - W^{(v)}(\phi_j)|} .\eea  Let $h^{(v)}$ be the real moment map in the complex structure $I^{(v)}$, explicitly \bea h^{(v)}  = v_1 \mu_I + v_2 \mu_J + v_3 \mu_K = \vec{v} \cdot \vec{\mu}.\eea  According to the Lemma, the real moment map $h^{(v)}$ is conserved along the gradient flow of $\text{Re}\big(e^{\text{i} \theta} W^{(v)} \big)$ (as is $\text{Im} \big(e^{\text{i} \theta} W^{(v)} \big)$) for any value of $\theta$. An $ij$ soliton can therefore only exist if \bea h^{(v)}(\phi_i) = h^{(v)}(\phi_j),\eea Letting $\vec{\mu}_i \in \mathbb{R}^3$ be the critical value of the hyperK\"{a}hler moment map in the $i$th vacuum, this condition is simply saying that \bea \vec{v} \cdot( \vec{\mu}_i - \vec{\mu}_j) = 0,\eea so that $\vec{v}$ is perpendicular to the vector \bea \vec{Z}_{ij} = \vec{\mu}_i - \vec{\mu}_j\eea pointing from the $j$th critical value to the $i$th critical value.  If this does not happen for any pair of distinct critical points, the soliton spectrum of the Landau-Ginzburg theory is empty. In the notation of \cite{Gaiotto:2015aoa}, the soliton space $R_{ij}$ is trivial \bea R_{ij} = \{0\},\eea for each pair of distinct critical points $(\phi_i, \phi_j)$. Since, in the Gaiotto-Moore-Witten formalism, morphism spaces between different thimble objects are given as direct sums of tensor products of soliton spaces, we obtain the following. 

\paragraph{Corollary} Suppose $\vec{v} \in S^2$ is such that $\vec{v} \cdot ( \vec{\mu}_i - \vec{\mu}_j) \neq 0$ for any pair of critical points $(\phi_i, \phi_j)$. Then the $A_{\infty}$-category of $\big((M, g, I^{(v)}), W^{(v)} \big)$ is semi-simple: \bea \text{Hom}(T_i, T_i) &=& \mathbb{C}, \text{ for all } i, \\ \text{Hom}(T_i, T_j) &=& 0 \,\,\,\,\text{ for } i \neq j.\eea

\paragraph{} One can also rephrase the discussion in terms Lefschetz thimbles and their intersections. The Lefschetz thimble $L_i(\theta;v)$ for a critical point $\phi_i$, and a given angle $\theta$, is the space of all gradient flow trajectories for the function $\text{Re}(e^{-\text{i}\theta}W^{(v)})$ that go to $\phi_i$ in the far past. We are simply saying that if $v$ is away from the exceptional locus, then the Lefschetz thimbles $L_{i}(\theta; v)$ and $L_{j}(\theta; v)$ for a given angle $\theta$, even when slightly rotated away from $\theta$ so that their images in the $W^{(v)}$-plane intersect, will have an empty intersection \bea L_i(\theta \pm \epsilon;s) \cap L_j(\theta \mp \epsilon; s) = \varnothing \eea in $M$.

\paragraph{} It is interesting to consider the exceptional loci, namely the points on the sphere of complex structures where $\vec{v} \cdot \vec{Z}_{ij} = 0$ for some pair $(i,j)$ of vacua. The exceptional locus is typically a set of great circles on the sphere, one such great circle $S_{ij}$ for each pair 
$(i,j)$ of vacua.  Along a point on the great circle $S_{ij}$ there can be an $ij$-soliton, and so a morphism space between distinct objects can be non-trivial.

\paragraph{} To get a better feeling for what happens at these exceptional points, we consider an example. Let $M$ be the cotangent bundle of the projective line $M = T^* \mathbb{P}^1$. Consider the complex structure induced from the complex structure on $\mathbb{P}^1$, and consider the holomorphic coordinates $(p,q)$ in one patch. We call this complex structure $I$. The real and holomorphic symplectic forms in this complex structure are given by \bea \omega &=& 2\text{i}\frac{\text{d}q \wedge \text{d} \bar{q}}{(1+ |q|^2)^2}  -2\text{i}(1+|q|^2)^2 \text{d}p \wedge \text{d}\bar{p}, \\ \Omega &=& \text{d}p \wedge \text{d}q. \eea The real and complex moment maps corresponding to the hyperK\"{a}hler isometry \bea (p,q) \rightarrow (e^{-\text{i}\theta} p, e^{\text{i} \theta} q),\eea are given by \bea h_I &=& \frac{1-|q|^2}{1+|q|^2} - (1+|q|^2)^2 |p|^2 \\ W_I&=& \text{i} pq.  \eea There are two critical points, $\phi_1$ and $\phi_2$ given by $(p,q) = (0,0)$ and $(p,q) = (0, \infty)$ respectively, with corresponding critical values \bea (h_I, W_I)|_{(0,0)}&=& (1,0), \\ (h_I, W_I)|_{(0,\infty)} &=&(-1,0). \eea In the complex structure $I$, the critical values of the real moment map are distinct, so that it satisfies the criteria of the Corollary. The Landau-Ginzburg superpotential in this complex structure, \bea W_I = \text{i} pq \eea has no flows between the distinct critical points, since the value of $W_I$ on both critical points is the same, being equal to zero. Therefore the $A_{\infty}$-category of the pair $ \big((T^*\mathbb{P}^1, g,I), W_I \big)$ is indeed semi-simple. On the other hand, consider the point on the sphere corresponding to the complex structure $K$. The holomorphic moment map in this complex structure is \bea W_K = h_I + \text{i} \text{Re}(W_I), \eea and the real moment map is \bea h_K = \text{Im}(W_I).\eea For the complex structure $K$ we indeed find that the real moment map $h_K$ has the same (vanishing) critical value for both critical points. The critical values of $W_K$ on the other hand are \bea W_K(\phi_{1}) = 1, \,\,\,\,\, W_K(\phi_2) = -1,\eea so the phase $e^{\text{i}\theta_{12}}$ of a BPS soliton interpolating between the critical points $\phi_1$ and $\phi_2$ must be real. We therefore study the gradient flow equation for $\text{Re}(W_K) = h_I$, or equivalently, the Hamiltonian flow equation for $\text{Re}(W_I)$ with respect to $\omega_K = \text{Im} (\Omega)$. This is the equation \bea \frac{\text{d}q}{\text{d}y} &=& -\text{i}\frac{\del W_I}{\del p}, \\ \frac{\text{d}p}{\text{d}y} &=& \text{i} \frac{\del W_I}{\del q},\eea which simply becomes  \bea \frac{\text{d}q}{\text{d}y} = q, \,\,\,\,\,\,\, \frac{\text{d}p}{\text{d}y} = -p.\eea There is a family of solutions: for each $q_0 \in \mathbb{C} \backslash \{0\}$ we have \bea q(y) &=& q_0 e^{y}, \\ p(y) &=& 0,\eea is a soliton interpolating between the critical points.  Writing $q_0 = e^{y_0 + \text{i} \alpha}$ we find that $y_0$ corresponds to an overall translation of the center, and $\alpha$ is the internal collective coordinate corresponding to the $U(1)$ isometry. Quantizing the latter collective coordinate, we conclude that the soliton space $R_{12}$ is isomorphic to the deRham cohomology of a circle.

\paragraph{} To give a little more detail on the latter point, note that the space of one-particle BPS states $\mathbb{M}_{12}$, namely the subspace of the Hilbert space $\mathcal{H}_{12}$ annihilated by the A-type supercharge (with $\zeta = 1$) and its adjoint is four-dimensional, since there are four fermion zero-modes: the superpartners $b,\ov{b}$ to the translational mode, and the superpartners $c, \bar{c}$ to the $U(1)$ global symmetry. We can also see these as coming from the fact that the BPS soliton equation is half-BPS in an eight-supercharge theory, giving us four broken supersymmetries which we identify as the fermion zero modes. These fermion zero modes act on the vacuum in $\mathcal{H}_{12}$ to generate an irreducible representation of a Clifford algebra with two creation and two annihilation operators \bea \{b, \ov{b} \} = \{c, \ov{c} \} = 1 ,\eea thus giving us the four-dimensional vector space $\mathbb{M}_{12}$. To obtain $R_{12}$, as done in \cite{Gaiotto:2015aoa}, we factor out the Clifford module corresponding to the translational mode $b$, so that \bea \mathbb{M}_{12} = \big(\mathbb{C}\oplus \mathbb{C}^{[1]} \big) \otimes R_{12} \eea leaving us with \bea R_{12} = \mathbb{C} \oplus \mathbb{C}^{[1]}.\eea 

\paragraph{} In a theory with two vacua $1$ and $2$, and $\zeta$ chosen so that $T_1< T_2$, the morphism space $$\widehat{R}_{12} := \text{Hom}(T_2, T_1),$$ coincides with the soliton space $R_{12}$. We therefore conclude that at the exceptional locus, the thimble objects $T_1, T_2$ satisfy \bea \widehat{R}_{12} = \mathbb{C} \oplus \mathbb{C}^{[1]}.\eea The $A_{\infty}$-structure on the category with objects $T_1$ and $T_2$ is the obvious one: the only non-trivial compositions come from the units in $\text{Hom}(T_1, T_1)$ and $\text{Hom}(T_2, T_2)$. 

\paragraph{} What happens at the complex structure $K$, and indeed for any $I_{(0,a,b)} = aJ + bK$ where $a^2 + b^2 = 1$, is that we are considering an A-model for a real symplectic form \bea \omega_{(0,a,b)} = \text{Im}\big( (a+ \text{i} b) \Omega \big),\eea that is exact (since $\Omega$ is an exact form on $T^* \mathbb{P}^1$). On the other hand, if we work with a complex structure $I_{(c,a,b)}$ with $c \neq 0$ (such as the complex structure $I$), then the real symplectic form $\omega_{(c,a,b)}$ is non-exact (since there's a non-zero contribution from the non-exact symplectic form $\omega$). The $A_{\infty}$-categories for exact and non-exact symplectic forms indeed behave differently. On the exceptional circle the symplectic manifold $(M, \omega_{(0,a,b)})$ is symplectomorphic to the real cotangent bundle $T^*S^2$, in which the non-trivial morphism space $\text{Hom}(T_2, T_1)$ was first worked out in \cite{Seidelpaper}.

\paragraph{} Another noteworthy feature of the exact locus is the following. Consider the zero-section of $ T^* \mathbb{P}^1$, which we denote as $C$. Note that $C$ is an $I$-holomorphic submanifold of $T^* \mathbb{P}^1$ such that the holomorphic symplectic form $\Omega$ vanishes when restricted to it. $C$ is therefore (the support of) what is known as a $(B,A,A)$ brane. In particular it is Lagrangian with respect to the symplectic structure $\omega_K$. We therefore expect it to be an object in the $A_{\infty}$-category of the pair $\big((T^*\mathbb{P}^1,g, K), W_K\big)$. Recall that a general object or \textit{brane} in the $A_{\infty}$-category generated by thimbles is given by a twisted complex. A twisted complex $B$ is specified by a choice of graded vector space $V_i$ for each thimble object $T_i$, along with a Maurer-Cartan element (also known as a boundary amplitude): an element \bea  \gamma_B \in \oplus_{i,j}V_i \otimes \widehat{R}_{ij} \otimes V^{\vee}_j \eea of degree $+1$ that solves the Maurer-Cartan equation. We write such a brane as \bea B = [\oplus_{i} V_i \otimes T_i, \gamma_B]. \eea We claim that the Lagrangian brane with support $C$ is equivalent to the following twisted complex: \bea B_C = [T_1^{[1]} \oplus T_2, \gamma_C]\eea where $T^{[1]}$ denotes a choice of a multiplicity space of $T$ being a one-dimensional vector space in degree $+1$ \bea T^{[1]} = \mathbb{C}^{[1]} \otimes T,\eea and $\gamma_C$ is a Maurer-Cartan element \bea \gamma_C \in \text{Hom}(T_1^{[1]} \oplus T_2,T_1^{[1]} \oplus T_2). \eea The degree one part of this space is one-dimensional and we let $\gamma_C$ be a spanning vector. Since \bea m_2(\gamma_C, \gamma_C) = 0,\eea it is a solution to the Maurer-Cartan equation. An elementary computation then shows that the cohomology of the endomorphism space of this object is \bea H^*\big(\text{Hom}(B_C, B_C) \big) = \mathbb{C} \oplus \mathbb{C}^{[2]}.\eea Moreover as an $A_{\infty}$-algebra this is nothing but the ring $\mathbb{C}[x]/x^2$ where $x$ is an element carrying homological degree $+2$. Thus the cohomology of $\text{Hom}(B_C, B_C)$ is the deRham cohomology ring of $C$, as we expect. We see that at an exceptional point the $A_{\infty}$-category admits a Lagrangian sphere as an object. 

\paragraph{} In summary, for $M= T^*\mathbb{P}^1$, the exceptional locus is a great circle on the twistor sphere corresponding to when the real symplectic form is exact. If we are away from this locus, the $A_{\infty}$-category of the holomorphic moment map is semi-simple, whereas on the exceptional locus there is a non-trivial morphism space isomorphic to the cohomology of a circle. It is also precisely at the exceptional locus that the $A_{\infty}$-category admits a Lagrangian sphere as an object. 

\paragraph{} Going back to the general case, at the exceptional locus, we see that the $A_{\infty}$-category depends on the spectrum of solitons for which in general there is no universal answer. However, there is one feature which we comment on: A BPS soliton in the hyperK\"{a}hler moment map setting is the critical point of a \underline{holomorphic} functional. Indeed, the gradient flow equation for $\mu_I$ is equivalent to both the $\omega_J$-flow for $\mu_K$ and the $\omega_K$-flow for $-\mu_J$. We can obtain the latter as the critical locus of the following holomorphic functional on the space of maps from $\mathbb{R}$ to $M$: \bea \mathcal{W}[\varphi] = \int \big(\varphi^*(\Lambda_c) + \text{i} \mu_c \,\text{d}y \big),\eea where $\varphi: \mathbb{R} \rightarrow M$, $\Lambda_{\text{c}}$ is the Liouville form for $\Omega_{\text{c}} = \omega_J + \text{i}\omega_K$ and $\mu_{\text{c}} = \mu_J + \text{i} \mu_K$. The functional $\mathcal{W}$ is indeed holomorphic in the complex structure on $\text{Map}(\mathbb{R}, M)$ induced from $I$, and a critical point of $\text{Re}\,\mathcal{W}$ obeys the $\omega_J$-flow equation for $\mu_K$. Thus we find that a soliton is now the critical point of a holomorphic functional. This gives another argment for why there are no instanton corrections even when there are non-trivial solitons. 

\paragraph{} To conclude, this note studies the $A_{\infty}$-category of a holomorphic moment map in some complex structure of a hyperK\"{a}hler manifold. Away from exceptional loci on the two-sphere of complex structures, the $A_{\infty}$-category is semi-simple. At the exceptional loci, while the category is not semi-simple in general, there are still no instanton corrections. We therefore find that the $A_{\infty}$-category of a holomorphic moment map is not very interesting. 

\paragraph{} Just like the $A_{\infty}$-category of a holomorphic function categorifies the MSW complex of its real part, it is natural to wonder if there is a categorification of the $A_{\infty}$-category of a holomorphic moment map that is richer, and more intrinsic to hyperK\"{a}hler moment maps. Such a proposal is supported by the fact that there is an uplift of the $\mathcal{N}=(4,4)$ theory to a three-dimensional theory with $\mathcal{N}=4$ supersymmetry, where the A-type supercharge lifts to a topological supercharge. This would suggest that the natural invariant associated to a hyperK\"{a}hler moment map is a suitable version of a 2-category.

\paragraph{} The conjectural 2-category associated to $(M, \vec{\mu})$ is currently under investigation.  

\section*{Acknowledgements} I thank Justin Hilburn, Mikhail Kapranov, Semon Rezchikov, Paul Seidel, and Edward Witten for stimulating discussions. This work is supported by the Institute for Advanced Study and the National Science Foundation under Grant No. PHY-2207584.

\end{document}